\documentclass{PoS}

\usepackage[T1]{fontenc} % if needed
\usepackage{amsmath}
\usepackage{slashed}
\usepackage{booktabs}
\usepackage{feynmp}
\usepackage{multirow}
\usepackage{graphicx}
\usepackage[section]{placeins}
\usepackage{longtable}
\usepackage{tabu}

\renewcommand{\imath}{\mathrm{i}}
\newcommand{\muu}[1]{\mu_u^{\text{eff,}#1}}

\title{A light singlet at the LHC and DM}

\ShortTitle{A light singlet at the LHC and DM}

\author{%\speaker
{Jan Kalinowski}\thanks{Dedicated to the memory of Maria Krawczyk}\\
        Faculty of Physics, University of Warsaw, ul.\ Pasteura 5, 02-093 Warsaw, Poland\\
        E-mail: \email{jan.kalinowski@fuw.edu.pl}}

\abstract{An interesting scenario of an R-symmetric supersymmetric model with a light singlet  is discussed. Since a light scalar in this model necessarily implies a light Dirac neutralino, its viability as a dark matter candidate is addressed. }

\FullConference{Corfu Summer Institute 2017 'School and Workshops on Elementary Particle Physics and Gravity'\\
		2-28 September 2017\\
		Corfu, Greece}

\begin{document}

\section{Motivation}
The N=1 supersymmetry algebra contains a single global U(1)$_R$ symmetry, called R-symmetry  \cite{Salam:1974xa,Fayet:1974pd}, with 
$[R,Q_\alpha]=-Q_\alpha$,  which implies that component fields of a superfield have R-charges 
differing by one unit.  R-symmetry is stronger than R-parity: since gauge vector fields carry no 
R-charge ($R=0$), gauginos have R-charge$=1$ and  therefore, 
unlike in the Minimal Supersymmetric Standard Model (MSSM),  Majorana gaugino mass terms in 
the soft SUSY breaking potential have $R=2$ and thus are inconsistent with R-symmetry. Gaugino 
mass terms of Dirac type can be introduced by enforcing an N=2 supersymmetric structure of the gauge/gaugino
sector, i.e.\  adding for  each gauge
group factor  a chiral superfield in the adjoint representation.  An exciting implication of R-symmetry 
 is that such a construction necessarily introduces additional scalar fields.  This is an interesting option since 
after the discovery of a Standard Model (SM)-like Higgs boson at the
LHC~\cite{Aad:2012tfa,Chatrchyan:2012xdj,atlascms}, it remains an open question whether 
there are additional
scalar particles, possibly even with smaller mass. 
Moreover, since R-symmetry forbids also Higgsino mass parameter $\mu$ and
all left-right sfermion mixings,  
alleviating  some CP- and flavour-violating constraints  \cite{Kribs:2012gx} and reducing 
production cross section for squarks  making squarks below
the TeV scale generically compatible with LHC data,  such a scenario is worth exploring.

\section{Introduction}
Supersymmetric extensions of the SM 
always predict additional scalars and scenarios 
with light scalars have been explored both in the minimal
supersymmetric standard model (MSSM) 
and in its extensions, such as
the NMSSM which contains a gauge singlet field.
While in the MSSM a light scalar is not particularly motivated, in  the NMSSM a light singlet-like scalar  pushes
 the tree-level value of the SM-like Higgs boson mass up towards  reducing fine-tuning 
 (see e.g.\ \cite{Bechtle:2016kui} and \cite{Ellwanger:2015uaz}  and references therein for the MSSM 
 and NMSSM, respectively).
It turns out that in the minimal R-symmetric supersymmetric model (MRSSM) a scenario with a light scalar is even more 
motivated since it is necessarily
connected with a very light neutralino  \cite{Diessner:2015iln} which can play a role of a dark matter particle.

The MRSSM is defined by the superpotential \cite{Kribs:2007ac}
\begin{eqnarray}
\nonumber W = && \mu_d\,\hat{R}_d \cdot \hat{H}_d\,+\mu_u\,\hat{R}_u\cdot\hat{H}_u\,+\Lambda_d\,\hat{R}_d\cdot \hat{T}\,\hat{H}_d\,+\Lambda_u\,\hat{R}_u\cdot\hat{T}\,\hat{H}_u\,\\ 
 && +\lambda_d\,\hat{S}\,\hat{R}_d\cdot\hat{H}_d\,+\lambda_u\,\hat{S}\,\hat{R}_u\cdot\hat{H}_u\,
 - Y_d \,\hat{d}\,\hat{q}\cdot\hat{H}_d\,- Y_e \,\hat{e}\,\hat{l}\cdot\hat{H}_d\, +Y_u\,\hat{u}\,\hat{q}\cdot\hat{H}_u\, ,
\label{eq:superpot}
 \end{eqnarray} 
where the MSSM-like fields are the Higgs doublet superfields $\hat{H}_{d,u}$
and the quark and lepton superfields $\hat{q}$, $\hat{u}$, $\hat{d}$,
$\hat{l}$, $\hat{e}$. 
 The new fields are the singlet, the SU(2)-triplet and the color-octet chiral 
superfields, $\hat{S}$, $\hat{T}$ and $\hat O$,  which contain the Dirac-mass
partners of the usual gauginos. Since the MSSM-like Higgs $\hat{H}_{d,u}$ are assumed to have $R=0$ 
in the MRSSM,  new  doublets $\hat{R}_{d,u}$ with $R=2$ are introduced (so called R-Higgs), which contain
the Dirac-mass partners of the higgsinos, and the corresponding Dirac-higgsino mass parameters 
$\mu_{d,u}$. In addition to the standard Yukawa couplings the superpotential contains 
also Yukawa-like
trilinear terms involving the new fields with 
 $\lambda_{d,u}$  parameters for the terms involving the singlet $\hat S$
and $\Lambda_{d,u}$ for the terms involving the triplet $\hat T$; terms involving the octet $\hat O$ are 
forbidden by R-symmetry.  Other parameters of the MRSSM   are the Dirac mass
parameters  $M_{B,W,O}^D$ for the U(1), SU(2)$_L$ and SU(3)$_c$  
 MSSM-like gauginos $\tilde B$, $\tilde W^a$ and $\tilde g^a$, which are paired with fermionic 
 components of $\hat S$, $\hat T$ and $\hat O$  respectively, 
the soft
scalar mass parameters $m^2_{S,T,O}$ and $m^2_{H_d,H_u,R_d,R_u}$ for the singlet, triplet, octet and  
 for the Higgs and R-Higgs states,  and the standard 
$B_\mu$ parameter and sfermion mass parameters, while the trilinear
sfermion couplings are forbidden by R-symmetry. For the explicit form of
the soft SUSY breaking potential see e.g.\ Ref.~\cite{Diessner:2014ksa}, and  for the detailed discussion of numerical tools used for results presented in this report  we refer to \cite{Diessner:2015iln}.

%%%%%%%%%%%%%%%%%%%%%%%%%%%%%%%%%%%%%%%%%%%%%%
\section{The MRSSM with a light singlet}

In the MRSSM
the  lightest Higgs boson tree-level mass is typically reduced
compared to the MSSM due to mixing with the additional scalars and
the loop corrections cannot be enhanced by stop mixing. However,
 the new fields and couplings give rise to the necessary large loop
contributions to the Higgs mass without generating too large a
contribution to the W-boson mass, as shown in Refs.~\cite{Diessner:2014ksa,Diessner:2015yna}
where  the
lightest (SM-like) Higgs boson mass has been computed at the one-loop
and leading two-loop level.  If on the other hand the SM-like Higgs is assumed to be the second-lightest, 
the mixing with additional scalars rises its tree-level mass and thus lowers the necessary loop 
corrections, similarly to  what occurs in the NMSSM.  

However, unlike in the NMSSM, 
requiring the second-lightest scalar to be the SM-like imposes constraints on bino-singlino and 
higgsino masses and couplings. To see this qualitatively, let us consider the neutral Higgs sector of the MRSSM. 
The real components of neutral scalar
fields $\phi_d,\phi_u,\phi_S,\phi_T$ of the two 
MSSM-like Higgs doublets $H_{d,u}$ and the $N=2$ scalar superpartners of
the singlet-triplet  gauge fields $S$ and $T$ do not mix with the corresponding imaginary 
components for real 
 vacuum expectation values  $v_{d,u,S,T}$. Therefore the full 8x8  mass-squared 
matrix breaks into two 4x4 sub-matrices. 
In the imaginary component sector the MSSM-like states  do not mix with 
the singlet-triplet  states  and the mass-squared matrix breaks further 
into two 2x2 sub-matrices. Thus the neutral Goldstone boson and
one of the pseudo-scalar Higgs bosons $A$ with $m_A^2=2B_\mu/\sin2\beta$  appear as in 
the MSSM. 
In the real component  sector all four states mix. However, 
the SM-like Higgs boson is dominantly given by the
up-type field $\phi_u$ for high $m_A$ and $\tan\beta=v_u/v_d$. In this limit it is enough to consider the 
2x2 sub-matrix of the neutral scalar mass matrix 
corresponding to the $(\phi_u,\phi_S)$ fields only, which reads 
\begin{eqnarray}
\mathcal{M}^{\phi}_{u,S}&= 
\begin{pmatrix} 
m_Z^2 +\Delta m^2_{rad}& v_u \left(\sqrt{2} \lambda_u \muu{-} +g_1 M_B^D\right) \\
v_u \left(\sqrt{2} \lambda_u\muu{-} +g_1 M_B^D\right) \; &
4(M_B^D)^2+m_S^2+\frac{\lambda_u^2 v_u^2}{2} \;\\ 
\end{pmatrix}
\;,
\label{eq:hu-s-matrix}
\end{eqnarray} 
where the dominant radiative correction to the diagonal component $\phi_u$ is denoted by $\Delta 
m^2_{rad}$, $g_1$ is the U(1) gauge coupling  and  
$\muu{-}$ stands for 
\begin{eqnarray}
\muu{-}&
=\mu_u+\frac{\lambda_u v_S}{\sqrt2}
-\frac{\Lambda_u v_T}{2}.
\end{eqnarray}
The parameters $M_{B}^D$, $\mu_{u}$ and $\lambda_{u}$ appear
in the scalar potential due to  the usual D-term generation of Dirac mass terms for the MRSSM. 

From this approximation it is evident that the
light-singlet scenario, in which $m_{H_1}<m_{H_2}\approx 125$ GeV, can be achieved if 
the singlet-diagonal element in Eq.~(\ref{eq:hu-s-matrix}) is smaller than the doublet-diagonal one,  
and off-diagonal elements small 
enough to avoid tachyonic states. From this requirement one can draw  
the following hierarchy 
\begin{eqnarray}
m_S, M_B^D  < 
m_Z  < \mu_u \, ,\qquad
|\lambda_u| \ll  1 \, .
\label{eq:hierarchy}
\end{eqnarray}
It  implies that not only $m_S$ but also the bino-singlino mass parameter $M_B^D$ must be of order 
tens GeV, a feature which is 
unique to the MRSSM.

\begin{figure}[t]
\includegraphics{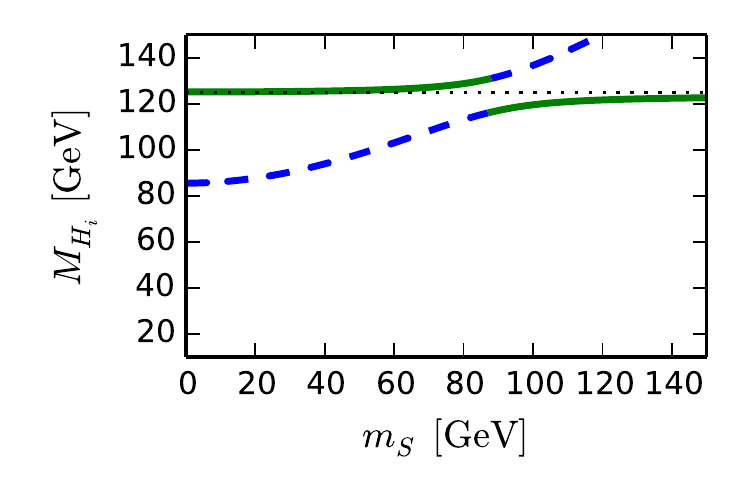}
\includegraphics{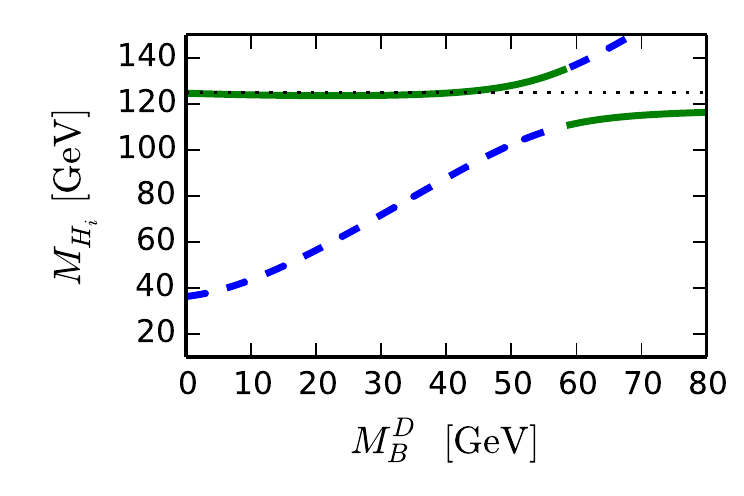}
\caption{Lightest and second-lightest CP-even Higgs states as functions of the dimensionful parameter $m_S$ ($M_B^D$) on the left (right); other parameters are as for BMP5 of Ref.~\cite{Diessner:2015iln}. 
The green full line indicates that the given state is more SM-like, while the blue dashed line when it is singlet-like.
}
\label{img:exclusion3}
\end{figure}

 A quantitative analysis of the masses 
of the two lightest Higgs states computed to two-loops  are shown in  Fig.~\ref{img:exclusion3} as 
a function of the two relevant
parameters $m_S$ (left panel) and $M_B^D$ (right panel). The non-varying parameters are fixed to the benchmark point BMP5 of
Ref.~\cite{Diessner:2015iln}, see Appendix for masses of some SUSY particles.  If the approximate
inequality Eq.~(\ref{eq:hierarchy}) is  satisfied, the
lightest state is significantly lighter than $m_Z$ and has a high
singlet component. When $m_S$ or $M_B^D$ become heavier, the lightest
state becomes mainly a doublet-like and hence SM-like.
In the light-singlet case, the upward shift of the SM-like Higgs can amount to more than
10~GeV, particularly due to the non-vanishing bino-singlino Dirac
mass $M_B^D$, since this parameter also appears in the off-diagonal
element of the mass matrix Eq.~\eqref{eq:hu-s-matrix}.

The novel feature of the light-singlet scenario in the MRSSM is
the upper bound on the Dirac bino-singlino mass $M_B^D$, which has
no counterpart e.g.\ in the NMSSM.
Obviously this bound affects the neutralino sector and with LEP bounds on chargino and neutralino production it
suggests that the LSP is a Dirac bino-singlino neutralino
with mass related to $M_B^D$ and thus limited from above.

%%%%%%%%%%%%%%%%%%%%%%%%%%%%%%%%%%%%%%%%%%%%%
\section{Light-singlet scenario at the LHC}

\begin{figure}[t]
\hspace*{1cm}\includegraphics[height=7.7cm]{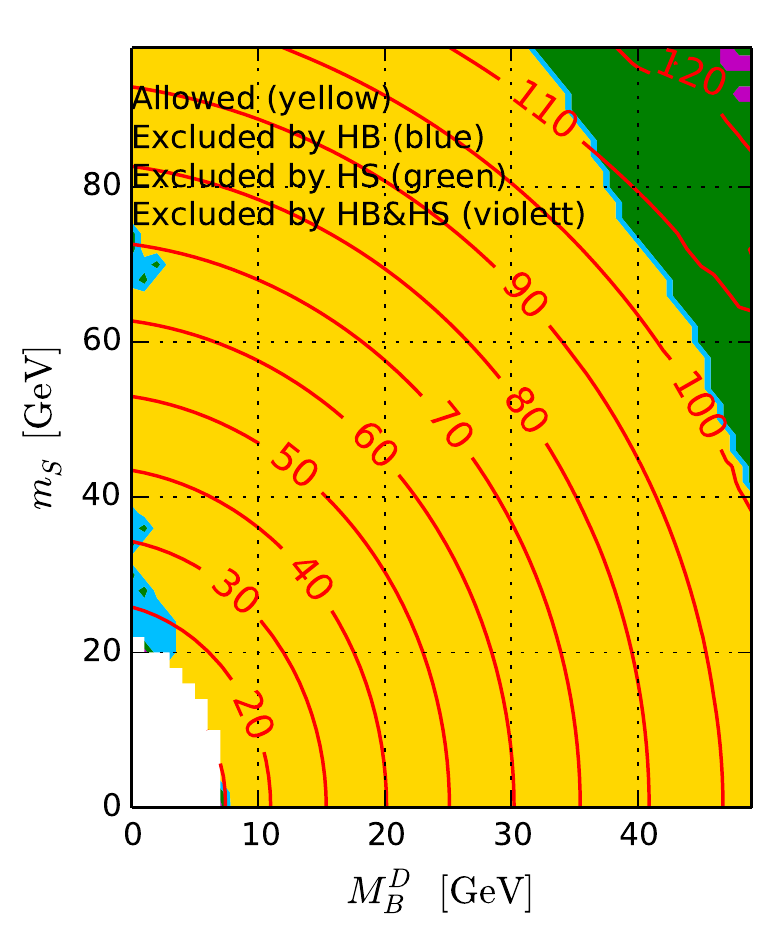}
\includegraphics[height=8cm,width=6.7cm]{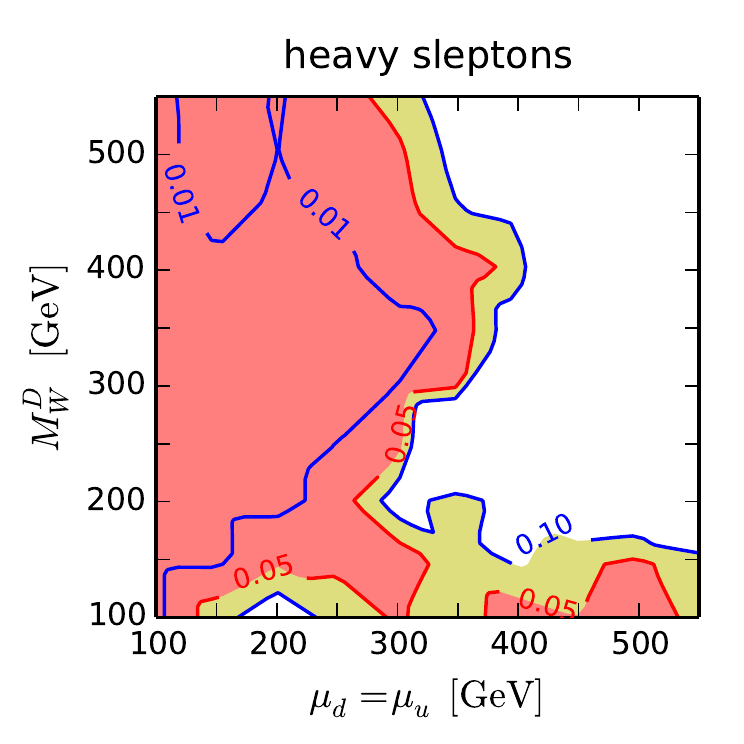}
\caption{Left: exclusion plot using \texttt{HiggsSignals-1.3} and \texttt{HiggsBounds-4.2} with $\Lambda_u$ 
scanned within the range $-1.5<\Lambda_u<0$.  
Red contours show the lightest scalar mass calculated at two loops.
Right: exclusion limits  for for heavy sleptons as a function
of the two higgsino mass parameters $\mu_{d}=\mu_u$ and wino-triplino mass $M^D_W$. 
The red (yellow) region marks the 95\% (90\%) excluded parameter. 
All other parameters
are fixed to the values of BMP5. }
\label{img:exclusion1}
\end{figure}

Experimental data impose  direct constraints on the two lightest scalars
and their mixing. First, there must be a SM-like state  observed at the LHC, with mass and couplings 
 to agree with the observed Higgs signal strengths and branching
ratios. Second, the  state lighter  than the SM-like 
one has to be mostly singlet-like  to 
pass limits from direct searches for light scalars, especially the LEP searches. 
Both constraints have been analyzed with 
\texttt{HiggsBounds}-4.2 and \texttt{HiggsSignals}-1.4 \cite{HB1} and 
the resulting excluded and allowed regions
in the plane of $m_S$ and $M_B^D$ are shown in the left panel of Fig.~\ref{img:exclusion1}.  
The parameter $\Lambda_u$ is adjusted to ensure the
correct mass of the observed Higgs boson; the remaining parameters are
fixed as in BMP5 with $\lambda_u=-0.01$. 
We find that the light-singlet scenario can be realized in the MRSSM
for $m_S<100$~GeV and $M_B^D<55$~GeV.

A light bino-like neutralino, and more generally light weakly
interacting particles, are interesting since they  might lead
to observable signals at the LHC and explain the observed
dark matter relic density. On the other hand, reinterpreting the LHC bounds in the context of the
MRSSM has to be done with care, since the MRSSM differs from the MSSM in the number of degrees
of freedom, different mixing patterns, Dirac-versus-Majorana nature of neutralinos, 
constraints due to the conserved R-charge etc.

In the slepton sector the left-right mixing vanishes in the MRSSM. 
This is not an essential difference to the MSSM, where typically  the mixing is
assumed to be small. Since in both models light sleptons 
decay to  leptons and bino-like
neutralino LSP, the Dirac or Majorana nature of the LSP is inessential and   
 the MSSM exclusion bounds for light sleptons can be applied  to the MRSSM case as well.  
The chargino and neutralino sectors  are
quite different. In the MRSSM, the four neutralinos are Dirac fermions
composed of eight Weyl spinors
$ ({\tilde{B}}, \tilde{W}^0, \tilde{R}_d^0, \tilde{R}_u^0)$
and 
$(\tilde{S}, \tilde{T}^0, \tilde{H}_d^0, \tilde{H}_u^0) $  with 
mass eigenvalues dominantly given by the four independent
mass parameters $(M_B^D,M_W^D,\mu_d,\mu_u)$. There are also four different charginos, 
with masses determined
by the wino and the two higgsino mass parameters. In the MSSM there are only two charginos and 
the neutralinos are Majorana with only a single
higgsino mass parameter $\mu$, so that two neutralino masses are
approximately degenerate. 
Further differences between MRSSM and MSSM exist in the
couplings of charginos and neutralinos and therefore in the decay
branching ratios.   Fixing the bino-singlino to be very light, $\sim50$ GeV, there are three
relevant mass parameters: the Dirac wino-triplino mass $M_W^D$ and the two
higgsino masses $\mu_{d,u}$. Further, the exclusion limits depend on
the slepton masses; 
for the electroweak and dark matter phenomenology all strongly interacting particles are 
not relevant  and are assumed to be heavy enough to evade limits.

As an illustration we consider a specific case with heavy sleptons which is of
interest for dark matter in the light-singlet scenario; for other parameter choices we 
refer to \cite{Diessner:2015iln}.  
Right panel of  Fig.~\ref{img:exclusion1} shows the exclusion regions
 in the
$\mu_d$=$\mu_u - M_W^D$ plane. Once both higgsinos are heavier 
than around 300~GeV there is a region with $M^D_W\geq 200$ GeV that is not excluded. 
The benchmark point MBP5 lies in this  viable parameter range.

\begin{figure}[t]
\hspace*{1cm}  \includegraphics[width=6cm, height=7cm]{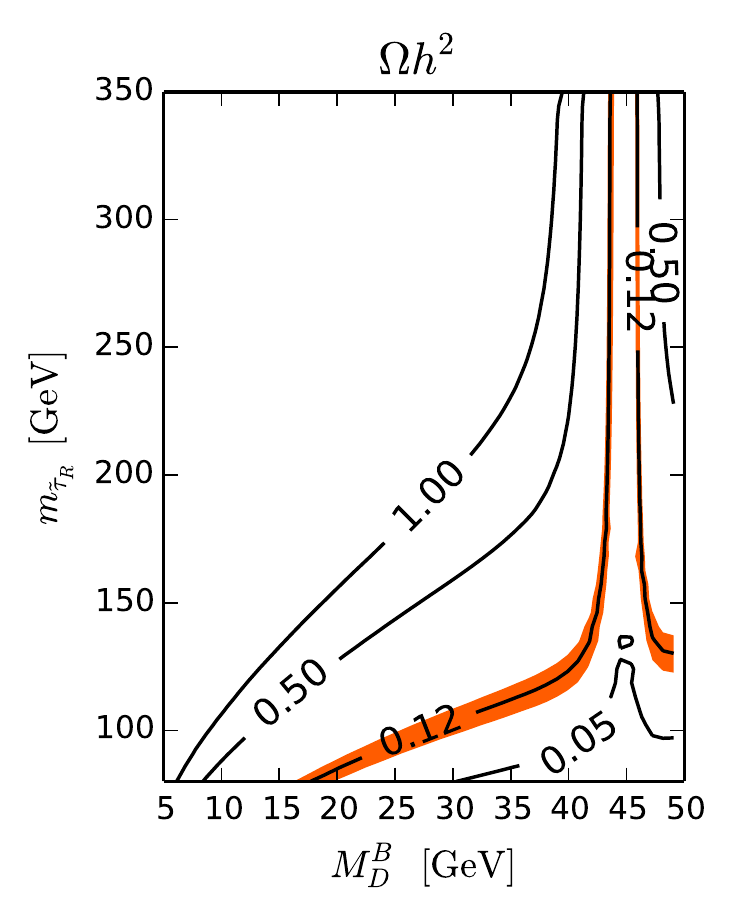}
  \includegraphics[width=6cm, height=7cm]{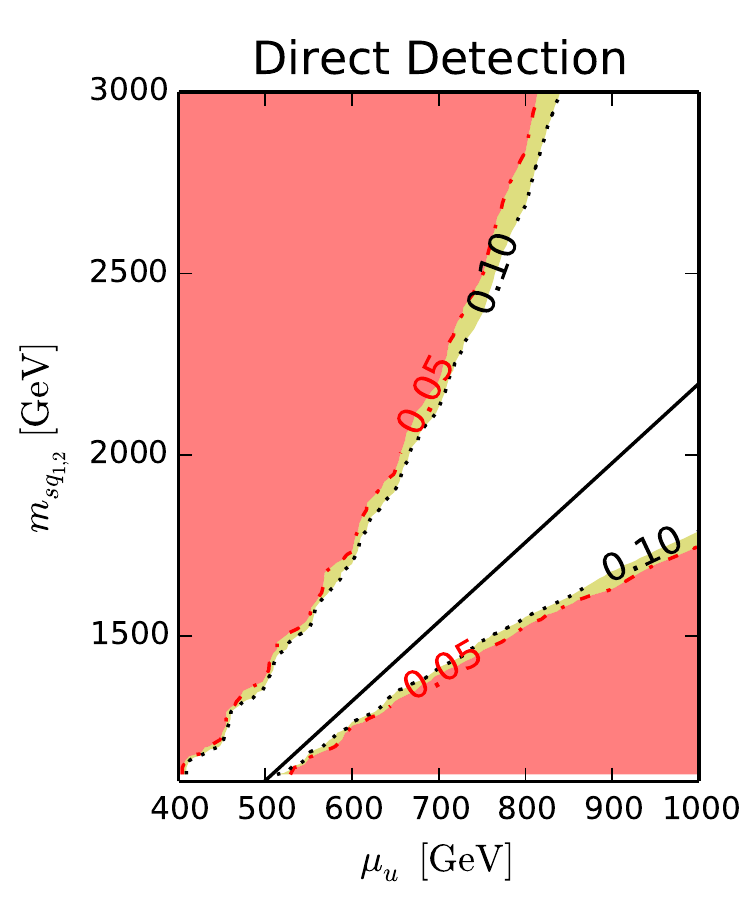}
  \caption{Left panel: dark matter relic density as a function of $M_D^B$  and $m_{\tilde \tau_R}$. 
Right panel: exclusion limits from direct detection  dending on $\mu_u$ and  equal first and 
second generation squark masses; the straight line shows the relation for 
the full destructive interference in Eq.~\eqref{eq:dd_not_simple_mass_rel}. 
}
\label{fig:dm}
\end{figure}

%%%%%%%%%%%%%%%%%%%%%%%%%%%%%%%%%%%%%%%%%%%%
\section{Light-singlet scenario and dark matter}
\label{sec:dm}
Demanding a light singlet Higgs state  leads to an upper bound on the bino-singlino mass 
parameter $M_B^D<55$~GeV. In this case 
the lightest neutralino is bino-singlino dominated, becomes the LSP and it is
the dark matter candidate of our model.  

For the DM relic density the crucial requirement,  as in the MSSM, is to achieve sufficiently
effective LSP pair annihilation processes. It turns out that two cases are
promising: (a) $m_{\chi_1}\approx M_Z/2$  and S-channel resonant LSP pair annihilation into Z bosons is
possible, or (b)  right-handed staus are light and annihilation
via t-channel stau exchange into tau leptons dominates.

Left panel of Fig.~\ref{fig:dm} shows the  allowed contour in the
$M_B^D$--$m_{\tilde{\tau}_R}$ parameter space. It clearly shows that the measured value 
of the relic density  can be met in  two different ways mentioned
above: at the $Z$ resonance or away from it. 
A sharp resonance-like peak is 
around $M_B^D\approx M_Z/2$ which results from the S-channel annihilation process, where 
the 
required stau mass has to be rather high. This is realized in the 
benchmark point BMP5. 
For $M_B^D$ away from the resonance the required stau
mass is below 150~GeV, as  in benchmark
points BMP4 and BMP6 of  \cite{Diessner:2015iln}.

For the direct DM searches, the spin-independent DM--nucleon scattering cross section at zero
momentum transfer can be written   in
terms of two scattering amplitudes $f_p,f_n$ as 
\begin{equation}
  \sigma_{DM-N} = \frac{4 \mu^2_{Z_A}}{\pi}\left(Z f_p  + (A-Z)f_n\right)^2\,.
\label{eq:dm-nuc-xs}
\end{equation}
Here $\mu^2_{Z_A}$ is the dark matter--nucleon reduced mass, and $A$
and $Z$ are atomic mass and number, respectively.
It has been  noted in Ref.~\cite{Buckley:2013sca} that the spin-independent
cross-section for Dirac neutralinos 
is dominated by the vector part of the Z boson-exchange and
squark-exchange contributions which  can lead to large scattering rates and thus to strong bounds on
the parameter space.

Since  the Z boson only couples to the
(R-)higgsino content of the LSP the corresponding amplitude can be  suppressed if  the higgsino mass parameters
$\mu_u$ and $\mu_d$ are sufficiently large.   
Similarly, for the squark-mediated amplitudes squarks  need to
be also sufficiently heavy.

The right panel of Fig.~\ref{fig:dm} shows the 95~\% and 90~\% exclusion bounds (violet (dark) and yellow (light dark) regions) 
derived using the log likelihood for the direct detection
by LUX 
depending on $\mu_u$ and the first/second generation squark masses.
The derived limits are quite sensitive to the
combination of both parameters. The funnel-shaped allowed region can be qualitatively explained by noting 
 a complete destructive interference between the Z- and 
squark-exchange contributions in the limit of degenerate heavy squarks and heavy higgsinos, i.e. $\sigma_{DM-N}=0$, when
  \begin{equation}
m_{\tilde{q}}^2= \frac{1}{3}(7+11 \frac{A}{Z-A})\mu_u^2 \phantom{o} \overset{\text{Xe}}{\approx} 
\phantom{o} 4.8\mu_u^2\;,
\label{eq:dd_not_simple_mass_rel}
\end{equation}
where the numerical value is for Xenon  with  $A=54$ and $Z=131.3$. 
The straight line in Fig.~\ref{fig:dm} (right)  corresponds to
Eq.~\eqref{eq:dd_not_simple_mass_rel}. Away from the line either the squark-mediated or  the Z-mediated
amplitude becomes small, the destructive interference does not work and experimental limit is not met. 
Note that the result for the exclusion bounds is calculated using the complete 
information of \texttt{micrOMEGAs}~\cite{Belanger:2014vza} and \texttt{LUXCalc}~\cite{Savage:2015xta},
while Eq.~\eqref{eq:dd_not_simple_mass_rel} is only approximated.
As the squark masses of the first two generations are limited by LHC searches, the direct detection
non-discovery provides a lower limit on $\mu_u$.

\begin{figure}[th!]
\includegraphics[height=6cm]{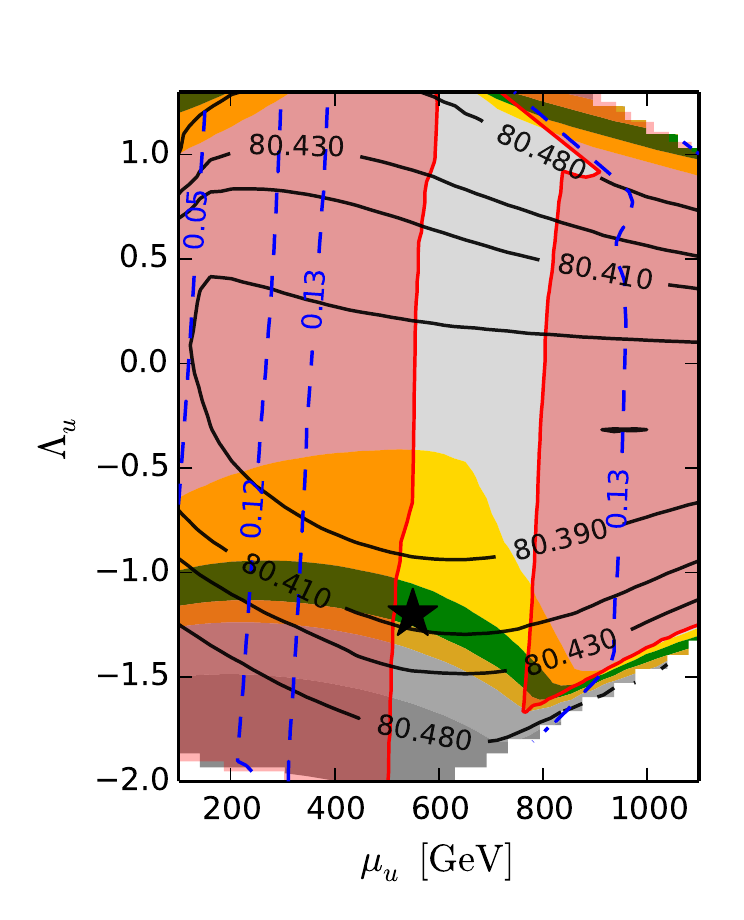}
\includegraphics[height=6cm]{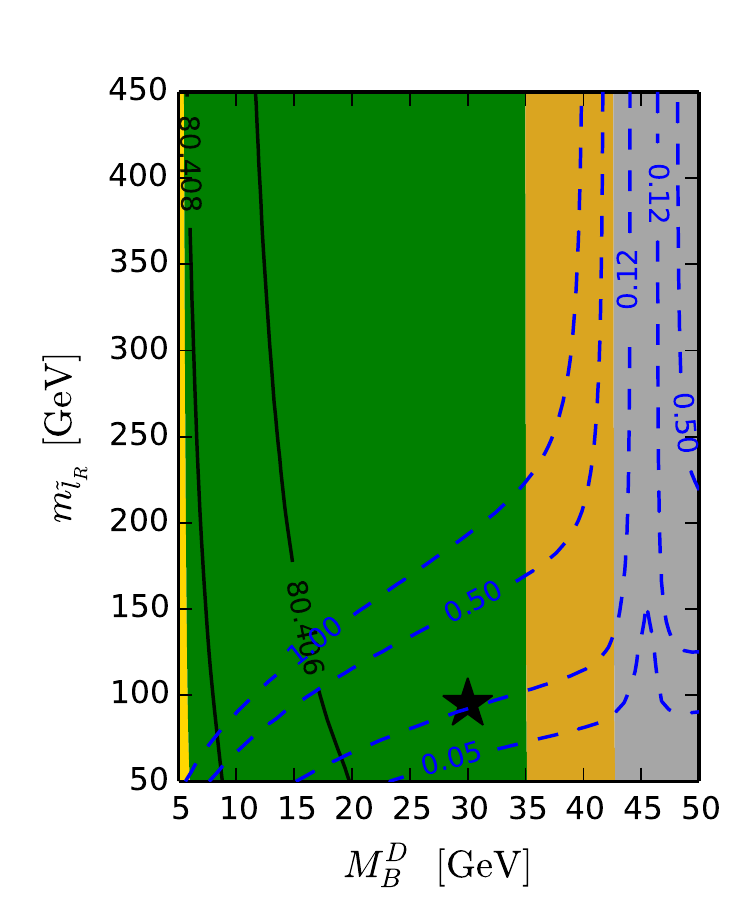}
\includegraphics[height=6cm]{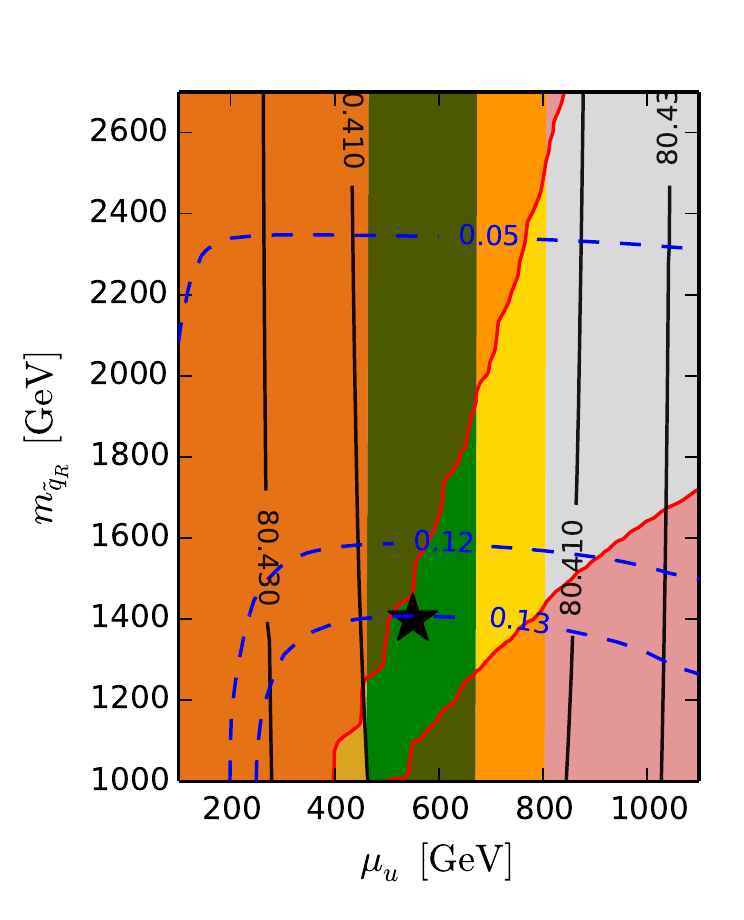}\\
 \includegraphics[width=12cm]{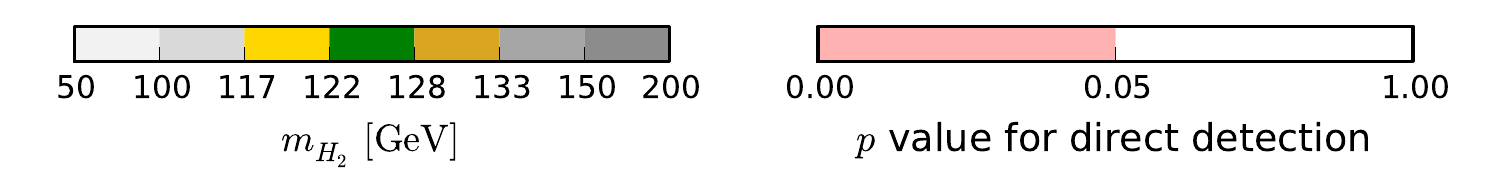}
\caption{Parameter regions contrasted  with experimental constraints.
Red  areas are  excluded with 95\% CL by dark matter direct detection, 
black full (blue dashed) lines show  $m_W$ ($\Omega h^2$). The mass of the
SM-like Higgs boson $m_{H_2}$ is given by the colour scale shown. All non-varied parameters are
set to the values of BMP6 and  BMP6 position is shown by stars.
}
\label{img:2dplots1}
\end{figure}

\section{Summary and outlook}
The minimal R-symmetric model MRSSM is a promising alternative to the
MSSM. In particular,  the light-singlet scenario is very attractive since it 
provides  an increased tree-level
SM-like Higgs boson mass and  many light weakly interacting particles
which could be discovered at the next LHC run as well as the possibility to
explain dark matter. It is also very predictive as can be inferred from  Fig.~\ref{img:2dplots1}.  
If a light singlet-like Higgs boson is found, 
not only the parameter $m_S$ but also $M_B^D$ and $\lambda_u$ must
be very small. Then, the requirement of  the 125 GeV Higgs mass essentially fixes $\Lambda_u$ and
constrains $M_B^D$ and $\mu_u$ (left panel), implying that the fermionic partner is the LSP. On the other hand dark matter 
data constrain slepton and squark masses (middle and right panels) and other electroweak 
parameters pointing to interesting mass hierarchies of SUSY particles to be searched at the LHC. It would also be interesting to explore  the ILC discovery potential at its initial 250 GeV phase \cite{Asai:2017pwp} for  the benchmark points discussed here.  

\section*{Acknowledgments}
I thank Philip Diessner, Wojtek Kotlarski and Dominik St\"ockinger for many valuable discussions and fruitful collaboration. 
Thanks also go to Gudrid Moortgat-Pick for her hospitality during my stay at DESY. Work supported in part by the Polish National Science Centre HARMONIA 
grant under contract UMO-2015/18/M/ST2/00518 (2016-2019) and the DFG through the SFB~676 ``Particles, Strings and the Early Universe''.

\section*{Appendix}
Masses of some SUSY particles relevant for the discussion in this report. All values are given in GeV. The four charginos appearing in the MRSSM are denoted as $\chi^\pm_1, \chi^\pm_2, \rho^\pm_1, \rho^\pm_2$. 
\begin{table}[ht!]
\centering
  \begin{tabular}{cccccccccccccc}
  \toprule
   & $\chi^0_1$ & $\chi^0_2$ &  $\chi^0_3$ & $\chi^0_4$ & $\chi^\pm_1$ & $\chi^\pm_2$ & $\rho^\pm_1$ & $\rho^\pm_2$& $\tilde{\tau}_R$& $\tilde{\mu}_R$& $\tilde{e}_R$ & $\tilde{\ell}_L$ & $m_{H_1}$\\
  \midrule
  BMP4& 49.8 & 132 & 617 & 691 & 131 & 625 & 614 & 713 & 128 & 802 & 802 & 808 & 100 \\
  BMP5& 43.9 & 401 & 519 & 589 & 409 & 524 & 519 & 610 & 1000 & 1001 & 1001 & 1005 & 94\\
  BMP6& 29.7 & 427 & 562 & 579 & 422 & 562 & 433 & 587 & 106 & 353 & 353 & 508 & 95\\
   \bottomrule
  \end{tabular}

%\caption{Masses of the non-SM particles in the BMPs  discussed here. All values given in GeV.  \label{tab:EW_mass}}
\end{table}

\end{document}